# Polarization eigenstates of axially symmetric waveguide modes in a uniaxial fiber

**Peter Muys**, Lambda Research Optics Europe

Tulpenstraat 2, B-9810 Eke-Nazareth, Belgium

Abstract

It was recently shown [1] that the polarization eigenstates of freely propagating axially symmetric modes along the c-axis of a uniaxial, dielectric crystal, exhibit radial and azimuthal polarization .

In this paper, the preceding results are extended from free wave propagation to guided wave propagation in a uniaxial, dielectric, circular fiber. The correct allocation of the different modes (TE vs. TM) to the corresponding propagation characteristics (ordinary vs. extra-ordinary) and to the corresponding polarization states (radial vs. azimuthal) will be disclosed.

The results are further used to define a new type of photonic crystal fiber, relying on form birefringence to impose the anisotropy. Since this type of fiber is naturally polarization maintaining, without the need to recourse to stress birefringence, as is required in regular fibers, it can be used as beam delivery system to transport radial/azimuthal polarization over large distances, e.g. for material processing applications, from the laser source to the point of impact.

## 1  Introduction

We will consider the electric field as a vector wave propagation in a lossless dielectric medium. The angular spectral wave expansion and the normal mode expansion are two classical methods to solve the vector wave equation in such a medium [1]. The first is most suited to treat free wave propagation phenomena, while the second is preferred when dealing with guided wave phenomena. Both methods also differ by how the boundary conditions are taken into account. As a consequence, this leads to a continuous or a discrete spectrum of the modes.

To study electromagnetic wave-guiding along the optical axis in a uniaxial c-cut crystal, the regular procedure to solve the vector wave equation is to start with the normal mode expansion. Here, the electric or magnetic field components are not any longer the familiar Cartesian components, but an equivalent decomposition in longitudinal (i.e. along the propagation axis in the crystal) and transversal components, called normal modes. However, the search for ordinary and extraordinary field propagation , as imposed by the birefringence of the crystal, is obscured heavily by the computational complexity as a consequence of the wave-guiding geometry. The major bottleneck is that the vector wave equation must be expressed in cylindrical coordinates, which in particular for the determination of the analytic form of the dielectric tensor is a long and tedious exercise.

In this paper, a hybrid approach is adopted to work around the above mentioned bottleneck. The approach is new, in the sense that it combines the methodology of the normal mode expansion with the methodology of the angular plane wave expansion. Its implementation is based on considering three consecutive steps.

Step 1

The angular plane wave spectrum of the electric field vector can be resolved in ordinary and extraordinary field components –this is normal- but *simultaneously* in longitudinal and transversal components, like in the normal mode approach[2]. This synchronism in decomposition is not evident and is crucial to our case. It effectively means that to study mode propagation in a crystal, a bridge can be laid from the angular spectrum representation (with its ordinary and extraordinary fields) to the normal mode expansion (with its transversal and longitudinal fields).

Step 2

The next step is the use of the theorem stating that, in c-cut uniaxial crystals, the ordinary field is possessing azimuthal polarization and the extraordinary field is assuming radial polarization. The proof of this theorem was published recently [3].

Step 3

The last step is the identification of the ordinary field as the lowest order TE mode of the dielectric waveguide with a circular cross-section.[Jackson]

The traditional paraxial approximation, introduced to simplify the computational mathematics, is not invoked and not required to obtain our results. The main assumption is that we consider the amplitude and the polarization of the (electric) fields to be axially symmetric, i.e. either radial or azimuthal. In fact, radial symmetry of the polarization is a sufficient condition for the radial symmetry of the field amplitude [4].

| normal mode | TE | TM |
|---|---|---|
| polarization | azimuthal | radial |
| propagation | ordinary | extra-ordinary |

Table 1: overview of the allocation

In summary, this paper describes the correct allocation of the polarization characteristics of the lowest mode propagating in a c-cut uniaxial fiber, respectively the ordinary mode versus the extraordinary mode, the TM-mode versus the TE-mode and the radial mode versus the azimuthal mode, see Table 1.

Since it is not required to derive explicit analytical expressions for the mode shapes in order to obtain our results, the field equations will actually not be solved. This is possible to do however, both analytically and numerically, but does not contribute to further insight into the physics of the problem.

## 2  the decomposition in an angular spectrum of plane waves

In this paragraph, the methodology described in detail in [3] and [4] is taken as starting point to solve the vectorial wave equation for waves propagating in an anisotropic dielectric. The main lines will be repeated here for convenience and will be adopted to the assumption of dealing with a c-cut uniaxial crystal where the optical propagation direction is parallel to the c-axis.

The z-axis coincides with the c-axis of the crystal, x and y are the transverse coordinate directions. The dielectric tensor of such a uniaxial crystal having its optical axis as crystallographic c-axis, is given in these Cartesian coordinates by

$$\tilde{\varepsilon} = \begin{bmatrix} n_o^2 & 0 & 0 \\ 0 & n_o^2 & 0 \\ 0 & 0 & n_e^2 \end{bmatrix} \qquad (1)$$

where $n_o$ and $n_e$ are the ordinary and extraordinary refractive indices respectively. The vector wave equation for the electrical field inside the crystal

$$\nabla^2 \mathbf{E} + k_0^2 \, \tilde{\varepsilon} \cdot \mathbf{E} = \text{grad div} \mathbf{E}$$

is solved by expressing the actual field $\mathbf{E}$ as a linear superposition of plane waves, taking into account the boundary value of the field at z=0. Here, $k_0$ is the vacuum wave number. These plane waves all are particular solutions of the wave equation. Their linear superposition is expressed mathematically as a two-dimensional Fourier transform. In the case of vectorial waves, it can be proved [4] that these plane waves decompose into two distinct categories, called resp. the ordinary waves and the extraordinary waves. If now the propagation is occurring along the z-axis, which is also the crystal c-axis, then the decomposition can be further made explicit. In this case effectively, the ordinary and extraordinary waves can be split *simultaneously* in their longitudinal (along the z-axis) and their transversal (perpendicular to the z-axis) components. First we recognize the beam shape character of the fields propagating along the z-axis:

$$\mathbf{E}(\mathbf{r}, z) = \exp(ik_0 n_0 z)\left[\mathbf{E}_o(\mathbf{r}, z) + \mathbf{E}_e(\mathbf{r}, z)\right] \qquad (2)$$

where the rapidly varying plane wave part along the z-axis has been factored out. Second, the component split in the right-hand side of Eq. 2 can be further resolved into transversal (T) and longitudinal (L) components as

$$\begin{aligned} \mathbf{E}_o(\mathbf{r}, z) &= \mathbf{T}_o(\mathbf{r}, z) + \mathbf{L}_o(\mathbf{r}, z) \\ \mathbf{E}_e(\mathbf{r}, z) &= \mathbf{T}_e(\mathbf{r}, z) + L_e(\mathbf{r}, z).\mathbf{e}_z \end{aligned} \qquad (3)$$

The corresponding ordinary and extraordinary propagation constants are defined by

$$\begin{aligned} \gamma_o &= \sqrt{k_0^2 n_o^2 - k^2} \\ \gamma_e &= \sqrt{k_0^2 n_e^2 - k^2} \end{aligned} \qquad (4)$$

where k is the transverse wave number, and equal in length to the length of the wave vector

$$\mathbf{k} = k_x \mathbf{e}_x + k_y \mathbf{e}_y$$

The ordinary and extraordinary transversal and longitudinal component of the electric field are now given for this specific case of a uniaxial crystal, by their angular spectrum of plane waves as:

$$\mathbf{T_o}(r,z) = \iint d^2\mathbf{k}\, \exp(i\mathbf{k}\cdot\mathbf{r} + i\gamma_o z)\mathbf{P_o}\cdot\mathcal{E}(\mathbf{k})$$
$$\mathbf{L_o}(r,z) = 0$$
$$\mathbf{T_e}(r,z) = \iint d^2\mathbf{k}\, \exp(i\mathbf{k}\cdot\mathbf{r} + i\frac{n_o}{n_e}\gamma_e z)\mathbf{P_e}\cdot\mathcal{E}(\mathbf{k}) \qquad (5)$$
$$\mathbf{L_e}(r,z) = \frac{-n_o}{n_e}\iint d^2\mathbf{k}\, \exp(i\mathbf{k}\cdot\mathbf{r} + i\frac{n_o}{n_e}\gamma_e z)\mathbf{k}\cdot\mathcal{E}(\mathbf{k})\frac{1}{\gamma_e}$$

The operators $\mathbf{P_o}$ and $\mathbf{P_e}$ are defined by the matrix expressions

$$\mathbf{P_o} = \frac{1}{k^2}\begin{pmatrix} k_y^2 & -k_x k_y \\ -k_x k_y & k_x^2 \end{pmatrix}$$
$$\mathbf{P_e} = \frac{1}{k^2}\begin{pmatrix} k_x^2 & k_x k_y \\ k_x k_y & k_y^2 \end{pmatrix} \qquad (6)$$

and can be identified as orthogonal projection operators. Also,

$$\mathcal{E}(\mathbf{k}) = \frac{1}{(2\pi)^2}\iint \mathbf{E}(\mathbf{r},0)\exp(-i\mathbf{k}\cdot\mathbf{r})d^2\mathbf{r} \qquad (7)$$

is the two-dimensional Fourier transform of the electrical field in the input plane z=0. It is important to note that, under the given assumptions, eq. (5) show that the ordinary electric field is purely transverse, and that the extraordinary field possesses a longitudinal component. This last one is normally neglected in free propagation studies, because it is so small. Neglecting this z-component is called the paraxial approximation.

As announced in the introduction, from here on, we suppose the polarization axially symmetric. This implies [1] that the fields will also be axially symmetric:

$$\mathbf{E}(\mathbf{r},0) = \mathbf{E}(r,0) \qquad (8)$$

We now treat first the consequences of this field symmetry. Next, we will introduce the consequences of the polarization symmetry.

So first, the two-dimensional plane wave spectrum of Eq. 7 can be reduced to a one-dimensional (with respect to the number of integration variables) zeroth order Hankel transform

$$\mathcal{E}(k) = \frac{1}{2\pi}\int_0^\infty \mathbf{E}(r,0)J_0(kr)r\,dr \qquad (9)$$

We see here that the angular spectrum inherits the polarization state of the as yet two-dimensional incident field.

We now state the influence of the orthogonal projection operators on the angular spectrum. The transverse ordinary and extraordinary field in the crystal and still for the general state of polarization, take on the following forms [4]:

$$\mathbf{T}_o(r,\phi,z) = \mathbf{A}_o^0(r,z) + \mathbf{R}(\phi) \cdot \mathbf{A}_o^2(r,z)$$
$$\mathbf{T}_e(r,\phi,z) = \mathbf{A}_e^0(r,z) - \mathbf{R}(\phi) \cdot \mathbf{A}_e^2(r,z) \tag{10}$$

where the matrix operator $\mathbf{R}$ is defined by

$$\mathbf{R}(\phi) = \begin{bmatrix} \cos(2\phi) & \sin(2\phi) \\ \sin(2\phi) & -\cos(2\phi) \end{bmatrix} \tag{11}$$

The respective field components in the right-hand side of Eq. 10 are

$$\mathbf{A}_o^n(r,z) = \pi \int_0^\infty \boldsymbol{\mathcal{E}}(k) \exp(iz\gamma_o) J_n(kr) k\, dk$$
$$\mathbf{A}_e^n(r,z) = \pi \int_0^\infty \boldsymbol{\mathcal{E}}(k) \exp(iz\frac{n_o}{n_e}\gamma_e) J_n(kr) k\, dk \qquad n=0,2 \tag{12}$$

Note how in Eq. 12 the respective $\mathbf{A}^n$ – vectors inherit the polarization state of the angular spectrum vector $\boldsymbol{\mathcal{E}}(k)$, according to:

$$\mathbf{A}_{o,e}^0 \parallel \boldsymbol{\mathcal{E}}(k)$$
$$\mathbf{A}_{o,e}^2 \parallel \mathbf{R}(\phi) \cdot \boldsymbol{\mathcal{E}}(k) \tag{13}$$

which, taking into account Eq. 10 for the polarization state of the ordinary and the extraordinary field, leads to

$$\mathbf{A}_o \parallel a\boldsymbol{\mathcal{E}}(k) + b\mathbf{R}(\phi) \cdot \boldsymbol{\mathcal{E}}(k)$$
$$\mathbf{A}_e \parallel c\boldsymbol{\mathcal{E}}(k) - d\mathbf{R}(\phi) \cdot \boldsymbol{\mathcal{E}}(k) \tag{14}$$

where a, b, c and d are (complicated) scalars defined through Eq. 12.

We now introduce the polarization symmetry of the fields. Eq 14 can be further reduced if the field polarization is assumed to be radial or azimuthal. We treat in parallel the case of radial and azimuthal polarization, implying for the incident field that

$$\mathbf{E}(r,0) = E(r,0)\mathbf{e}_r$$
$$\mathbf{E}(r,0) = E(r,0)\mathbf{e}_\phi \tag{15}$$

respectively. This immediately leads for the respective two-dimensional Fourier transforms to

$$\mathcal{E}(k) = \mathcal{E}(k)\mathbf{e}_r$$
$$\mathcal{E}(k) = \mathcal{E}(k)\mathbf{e}_\phi \quad (16)$$

The radial and azimuthal unit vectors are linked with their Cartesian counterparts by the orthogonal coordinate transformation

$$\mathbf{e}_r = \cos\phi \cdot \mathbf{e}_x + \sin\phi \cdot \mathbf{e}_y$$
$$\mathbf{e}_\phi = -\sin\phi \cdot \mathbf{e}_x + \cos\phi \cdot \mathbf{e}_y \quad (17)$$

Application of the matrix $\mathbf{R}(\phi)$ on these unit vectors, using Eq. 17, results in:

$$\mathbf{R}(\phi) \cdot \mathbf{e}_r = \mathbf{e}_r$$
$$\mathbf{R}(\phi) \cdot \mathbf{e}_\phi = -\mathbf{e}_\phi \quad (18)$$

We now identify the azimuthal polarization with the ordinary field component and the radial polarization with the extraordinary field component. In [2] we showed that this particular substitution is in agreement with the required field direction of the dielectric displacement field vector **D** inside the crystal. This means we substitute the second of Eq. 16 in the first of Eq. 14 for the ordinary field, and the first of Eq. 16 in the second of Eq. 14 for the extraordinary field. This translates Eq. 14 into

$$\mathbf{A}_o \parallel [a\mathcal{E}(k)\mathbf{e}_\phi + b\mathcal{E}(k)\mathbf{R}(\phi) \cdot \mathbf{e}_\phi]$$
$$\mathbf{A}_e \parallel [c\mathcal{E}(k)\mathbf{e}_r - d\mathcal{E}(k)\mathbf{R}(\phi) \cdot \mathbf{e}_r] \quad (19)$$

Finally, by substituting the relations Eq. 18 in Eq. 19, we arrive at

$$\mathbf{A}_o \parallel [a\mathcal{E}(k) - b\mathcal{E}(k)]\mathbf{e}_\phi$$
$$\mathbf{A}_e \parallel [c\mathcal{E}(k) - d\mathcal{E}(k)]\mathbf{e}_r \quad (20)$$

Hence, the transverse o-field is completely azimuthally polarized, and the transverse e-field is completely radially polarized.

To compare later in the paper with the normal mode expansion method, we summarize our findings using the angular spectral mode expansion for a c-cut uniaxial crystal up to now.

*The ordinary electric field vector is purely transverse and is azimuthally polarized. The extraordinary electric vector has longitudinal and transversal components and its transversal component is radially polarized.*

## 3  the normal mode expansion

Now , we consider propagation along a round dielectric fiber. For waveguide propagation, the boundary conditions of the fields at the fiber outer surface must be properly matched. For this situation, the normal mode expansion is more appropriate than the plane wave expansion. In the normal mode expansion, the electric and the magnetic field are considered simultaneously, because of the need to satisfy simultaneously the electric and magnetic boundary conditions at the  outer edge of the wave guide. Alternatively, in the plane wave expansion, only the electric field is considered, although the magnetic field is implicitly always present of course. There is no need to keep track of the magnetic field because for free propagation , no simultaneous boundary conditions need to be satisfied during propagation.

The boundary conditions in the normal mode expansion imply that only modes fulfilling a resonance condition will be guided. The mode spectrum  hence becomes discrete, but the normal mode set mathematically still exhibits  completeness [6].

Applying the normal mode expansion to a circular dielectric fiber having a refractive index distribution which is radially symmetric, leads to the identification of TE and TM modes. The $TE_{0p}$ and $TM_{0p}$ , p= 1,2, … , are the lowest order modes in such a circular isotropic dielectric fiber[5,6]. Radial symmetry of the refractive index  is effectively occurring around the c-axis of the uniaxial crystal. So the conditions for the application of the TE/TM decomposition theorem are fulfilled. For the TE mode, the longitudinal electric field component is zero, whereas $E_\varphi$, $H_r$ and $H_z$ are non-zero and inside the fiber are given by

$$H_z = aJ_0(\gamma_o r)$$
$$\mathbf{H}_t = H_r \mathbf{e}_r = \frac{ik_z}{\gamma_o^2} \nabla_t H_z \qquad (21)$$
$$\mathbf{E}_t = E_\phi \mathbf{e}_\phi = -\frac{i}{\gamma_o^2} \omega \mu_0 \mathbf{e}_z \times \nabla_t H_z$$

$\nabla_t$ is the transverse gradient operator. Note that we plugged in the ordinary propagation constant $\gamma_o^2 = k^2\omega^2/c^2 - k_z^2$ in these equations. *Eq. 21 show that the TE-mode is azimuthally polarized.  Similarly, the TM-mode is radially polarized.*

## 4  Merging the methods

Now follows the final step of our approach. We saw in paragraph 2 that the (non-paraxial) angular spectrum theory predicts that the ordinary field is azimuthally polarized. The modal theory of paragraph 3 on its turn shows that the TE-mode is azimuthally polarized. Both methods are mathematically equivalent so must give the same solutions. *Hence we conclude that the ordinary field can be identified as the lowest order TE mode. In a similar way, the extraordinary field is radially polarized and is the lowest order TM mode of the waveguide.*

## 5  Uniaxial photonic crystal fibers

The above theory could be checked experimentally for its validity, if a mono-crystalline fiber would be available. It is a quite challenging engineering task to draw mono-crystalline fibers however. As a possible compromise, we suggest that form birefringence of isotropic dielectric rods can be invoked to create an artificial crystal structure having nevertheless a fiber shape. To this end, consider the form birefringence occurring in the periodic piling of round dielectric rods, according to a square stacking pattern. Wiener [7] has shown that this sub-wavelength configuration effectively behaves as a uniaxial crystal. More specific, he considered dielectric rods imbedded in a vacuum environment. Here, we consider the complementary case of vacuum "rods" in an embedding dielectric. In other words, circular holes are present in a normal section of the dielectric fiber, patterned along a square grid. For this case, the Wiener effective medium model gives for the birefringence:

$$n_e^2 - n_o^2 = \frac{f_1 f_2 (1-n^2)^2}{(1+f_1)n^2 + f_2} \qquad (22)$$

where $f_1$ is the ratio of the diameter of the holes to their separation distance, $f_2 = f_1 - 1$, and n is the refractive index of the dielectric.

We hence arrive at a fiber structure showing waveguiding capabilities enabled by the presence of the regularly patterned air holes. It could therefore eventually be named a holey fiber too. In a regular holey fiber however, a central hole is missing. The effective medium theory explains the waveguiding by the contrast in optical density in the central area of the fiber, as compared to the peripheral zone of the fiber. Here, we invoked the same effective medium theory to show form birefringence of the fiber, which is now acting as a long uniaxial anisotropic waveguide. We dubbed it uniaxial photonic crystal fiber.

In summary, we showed that in a uniaxial, round, dielectric waveguide the lowest order ordinary wave is a TE wave exhibiting radial polarization.
We applied this theorem to define a new type of waveguide, showing form birefringence to make it behave like a uniaxial crystal, and which consequentially will support radial and azimuthal guided modes. This means the fiber is polarization maintaining for both radial and azimuthal polarization. It can find technological applications in the field of materials processing. Moreover, the dielectric can be doped with a proper impurity ion, so that the fiber can be used as active material in a fiber laser, which will then generate directly radial or azimuthal polarization.